\documentclass[a4paper]{article}
\usepackage[textwidth=30mm]{todonotes}
\usepackage{INTERSPEECH2020}
\usepackage{color, colortbl}
\definecolor{Gray}{gray}{0.9}
\usepackage{tabularx,booktabs}
  \newcolumntype{Y}{>{\centering\arraybackslash}X}

\usepackage{float}
\usepackage{amsmath}
\usepackage{amsfonts}
\usepackage{amssymb}

\restylefloat{table}

\title{Embedded Emotions - A Data Driven Approach to Learn Transferable Feature Representations from Raw Speech Input for Emotion Recognition}
\name{Dominik Schiller $^1$, Silvan Mertes $^1$, Elisabeth Andr\'e $^1$ }
\address{
$^1$Human Centered Multimedia, Augsburg University, Augsburg, Germany}
\email{schiller, mertes, andre@hcm-lab.de}

\begin{document}

\maketitle
\begin{abstract}
Traditional approaches to automatic emotion recognition are relying on the application of handcrafted features. 
More recently however the advent of deep learning enabled algorithms to learn meaningful representations of input data automatically.
In this paper, we investigate the applicability of transferring knowledge learned from large text and audio corpora to the task of automatic emotion recognition.
To evaluate the practicability of our approach, we are taking part in this year's Interspeech ComParE Elderly Emotion Sub-Challenge, where the goal is to classify spoken narratives of elderly people with respect to the emotion of the speaker. 
Our results show that the learned feature representations can be effectively applied for classifying emotions from spoken language. 
We found the performance of the features extracted from the audio signal to be not as consistent as those that have been extracted from the transcripts.
While the acoustic features achieved best in class results on the development set, when compared to the baseline systems, their performance dropped considerably on the test set of the challenge. 
The features extracted from the text form, however, are showing promising results on both sets and are outperforming the official baseline by 5.7 percentage points unweighted average recall.

\end{abstract}
\noindent\textbf{Index Terms}: emotion recognition, deep learning, representation learning, transfer learning, computational paralinguistics, natural language processing

\begin{figure*}[t]
 \centering
  \includegraphics[width=0.8\linewidth]{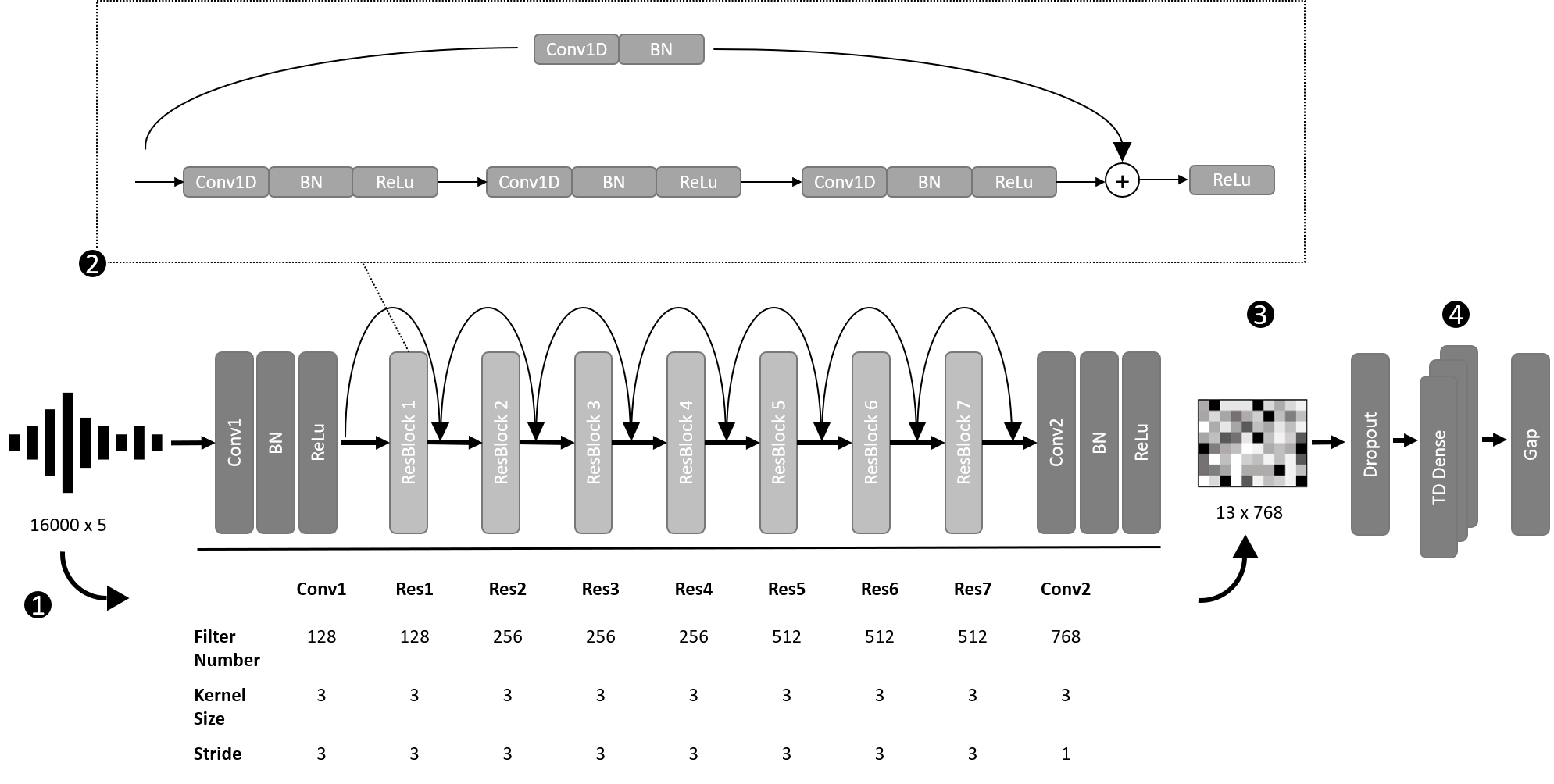}
  \caption{Architecture of the acoustic classification model. The raw 16kHz waveform is fed to the fully convolutional model in 5 second chunks (1). To train a deeper network with smaller filter kernels effectively, residual connections are added to each convolutional block (2). The outputs of the convolutional network are the final extracted feature representations. 
  During the training of the model those features are fed to a dense layer classification part (4).}
  \label{fig:system architecture acoustic}
\end{figure*}

\section{Introduction}
The challenging task of automatic emotion recognition is still subject of ongoing research.  
Emotional states are expressed in very individual ways among human beings, and marginal changes in the shades of these expressions can lead to a completely different perception of emotion. 
Due to the fine granularity of this problem domain, it is a very difficult and time consuming task to identify and craft features that allow the recognition of variables representing the emotional state, e.g. the valence and arousal values that occur during a spoken sentence. 
Traditional approaches rely on large sets of handcrafted representations, e.g. the \textit{ComParE} feature set for the acoustic domain \cite{weninger2013acoustics} or the creation of large lexica to map words to their emotional meaning for the textual domain \cite{mohammad2013crowdsourcing}.
The continuously ongoing success of deep learning has facilitated the automatic learning of features across many domains in a supervised and unsupervised fashion using convolutional neural networks (CNNs). 
Various research has shown the benefits of such a fully hidden feature extraction approach, commonly known as \textit{Representation Learning}, for recognizing emotions from audio \cite{wagner2018handcrafted, trigeorgis2016adieu} as well as textual input \cite{polignano2019comparison}. 

The idea behind this concept is to enable a machine learning system to derive feature embeddings directly from raw or low level input data without having feature engineers involved in the selection process. 
Such approaches are leading to a substantial decrease in time that has to be invested for crafting feature representations manually, while simultaneously overcoming performance limitations of the recognition system that result from human misjudgments during the feature crafting stage.
A drawback of those representation learning systems is that large amounts of training data are necessary to achieve state of the art performance. 
Specifically for the acoustic signal, current approaches often rely on learning the representations directly from the target corpus used for the final classification task \cite{wagner2018handcrafted, compare2020, rajan2019conflictnet, trigeorgis2016adieu}, which in return might result in suboptimal performance on small datasets. 

For tasks like automatic emotion recognition, where the labeling process has to be performed manually and is therefore extremely costly, labeled data is usually scarce, which presents a challenge for such data hungry algorithms \cite{park2019specaugment, perez2017effectiveness, fadaee2017data}.

However, there are recognition tasks in the same domains for which huge labeled datasets are available. 
The LibriSpeech dataset for example \cite{panayotov2015librispeech} consists of 1.000 hours of read English speech, labeled with features such as \textit{speaker identity} and \textit{gender}. 
Similarly, there is a plethora of data available that can be utilized to learn general and semantically meaningful representations for textual data, e.g. news articles or Wikipedia dumps. 
Studies \cite{sharif2014cnn, gwardys2014deep} have shown that CNNs are able to learn general features, that can be successfully used to solve various tasks which might differ significantly from the original training target of the model. 
Since most of those works are only studying this methodology with respect to computer vision problems, we wonder if the same holds true for text or raw audio input. 
Considering the lack of large datasets that are labeled with respect to a speakers emotion, we investigate whether such automatically learned representations from audio and textual input on large corpora can be successfully transferred to the task of automatic emotion recognition.

To evaluate the practicability of our approach, we participate in this year's Interspeech ComParE Elderly Emotion Sub-Challenge\cite{compare2020}.
The goal of this challenge is the recognition of the emotional state of elderly people in terms of discrete valence and arousal values from a German audio recording and the aligned transcript.
To assess the generalizability of our proposed approach regarding different domains, we decide to individually learn feature representations for audio and text respectively. 

For the audio domain we learn those feature representations from a gender classification task on the LibriSpeech dataset \cite{panayotov2015librispeech}. 
For the task of emotion recognition, current state-of-the-art approaches for representation learning of audio signals are mostly relying on spectrograms as inputs \cite{yenigalla2018speech, fedotov2019time, compare2020}, since those have achieved superior performance over raw waveform inputs.
In this work however we focus specifically on the application of raw waveforms as input, since every prior transformation of the input could result in a potential loss of information.
We hypothesize that an increased depth of the feature extraction CNN, made possible by the large amounts of available training data, can compensate for the inefficient representation of the signal domain. 

For the text domain we use an existing \textit{Bert Model} that is trained on a large dataset consisting of data extracted from Wikipedia, legal texts and news articles in an unsupervised fashion to learn semantically rich feature representation for a sentence de-masking task.
To evaluate the effectiveness of the extracted representations for the task at hand, while maintaining the comparability of our results, we are relying on the same evaluation approach as the authors of the challenge.
That is, we train a \textit{Support Vector Machine} (SVM) for valence and arousal values on the extracted feature embeddings. 

The remainder of this work is structured as follows:
In chapter \ref{approach} we introduce our approach in detail. 
Chapter \ref{datasets} briefly describes the dataset that was used in the challenge. 
We present our results and compare them to the challenge's baseline approaches in chapter \ref{results}, before we finally discuss our findings in chapter \ref{discussion} and draw some conclusions in chapter \ref{conclusion}.

\section{Methodology}\label{approach}
The study at hand approaches the question of whether generic, automatically learned features can be successfully applied to the task of emotion recognition from longer speech segments. 
It is well known that arousal can be better estimated by analyzing the acoustic properties of speech, while valence can be assessed more effectively through the semantic content of spoken utterances \cite{karadougan2012combining}.
To capture both dimensions optimally we employ two different neural network based feature learning systems.
The first system receives the raw audio waveform as input, in order to capture the paralinguistic emotional cues, while the second system analyzes the transcription with respect to the semantic content.
The following section provides an overview of the overall applied methodology and a detailed description of both feature extraction systems.

\subsection{Acoustic System}\label{sal}

The architecture of our acoustic classification network is closely modeled after the SampleCNN model, which was developed by Lee et al. \cite{lee2018samplecnn} for the task of music genre classification on the raw audio waveform.
The model is specifically designed to overcome the limitations of previous models, which have focused on simulating the behaviour of a frame-level time-frequency transformation (i.e. spectrograms) \cite{dieleman2014end}.
To this end the SampleCNN architecture relies on the application of many stacked 1D convolutional layers with very small filter sizes.
This essential design element also sets the model apart from other models that were used to detect emotions on the raw waveform \cite{trigeorgis2016adieu, wagner2018handcrafted}.
Recently Kim et al. \cite{kim2019comparison} showed that the SampleCNN architecture is capable of catching meaningful characteristics of sounds produced by the human vocal tract, when they applied the model to the task of keyword spotting. 
In their studies, the network achieves superior performance when compared to spectrogram based models. 

Our model consists of one, 1D convolutional layer in the beginning followed by a batch normalization layer and a relu activation layer. 
After that we are applying seven residual blocks with an increasing number of filters and a fixed stride of three to downsample the signal in each block. 
At the end of the feature extraction part we apply another convolutional layer with 768 filters and a stride of one followed by a batch normalization and a relu activation layer.
We use the output of this layer as features for our final classification task for the challenge.
However to learn meaningful representations we first need to fit the model to a pretraining task. 
To this end we are applying a dropout layer with a dropout factor of 0.5 before we add a multiple dense layers, distributed over time, to classify each time step of the output feature map using a softmax activation function.
The final output is calculated by averaging over the output of all dense layers.
The complete system architecture of the acoustic model is depicted in Fig. \ref{fig:system architecture acoustic}. 

To learn a suitable representation of the audio signal produced by the human voice we train our feature extraction model on the task of automatic gender recognition from raw audio on a large dataset. 
To this end we use the LibriSpeech dataset \cite{panayotov2015librispeech} which consists of 1.000 hours of speech extracted from English audio book recordings, distributed over \texttildelow 270.000 samples with an average of 12 seconds. 
We want to point out that the target language of our final model is German, which differs from the source language of the LibriSpeech dataset. 
As a result we expect that our final emotion classification model will focus on the acoustic properties of the voice while ignoring any potentially learned language specific semantics (e.g. keywords).

For training, all input data has been resampled to 16kHz mono channel audio.

For each sample we extract 5 seconds of audio randomly and apply data augmentation and preprocessing during training to increase the robustness of the feature extraction model and avoid overfitting.
Specifically, we apply a random shift along the time axis of up to 20\% of the total input length and add small amounts of both uniformly distributed and normal distributed random noise.
All input data has been locally normalized to have zero mean. 

To train the network we use the Adam optimizer with a learning rate of 0.001, a batch size of 32 and categorical cross entropy as a loss function. 
We train the model for 10 epochs. 
The final model achieves a performance of 97\% accuracy on the official test partition of the dataset.

To calculate the final feature vectors for training we concatenate the mean and the maximum value for each extracted filter response per sample, which results in 1536 distinct features. 

 \begin{table*}[t!]
\centering
\caption{Results on the development set for end-to-end and low-level descriptors with respect to the number of training epochs.}
\label{tab:results-dev}
\scriptsize 
\begin{tabularx}{\linewidth}{Y|*{2}{Y}|*{2}{Y}|*{2}{Y}|*{2}{Y}}
\toprule

\multicolumn{1}{c}{} & \multicolumn{4}{c|}{Arousal UAR[\%] Dev} & \multicolumn{4}{c}{Valence UAR[\%] Dev}\\ \midrule
\multicolumn{1}{c}{} & \multicolumn{2}{c}{Proposed} & \multicolumn{2}{c|}{Baseline} & \multicolumn{2}{c}{Proposed} & \multicolumn{2}{c}{Baseline} \\\midrule
\multicolumn{1}{c|}{C} & SampleCNN & SentenceBert & DeepSpectrum & LiFE & SampleCNN & SentenceBert & DeepSpectrum & LiFE \\ \midrule        

10$^{-5}$ & 44.6 & 39.4 & 37.4 & 33.3  & 33.3 & 33.3 & 33.3 & 33.3 \\
10$^{-4}$ & 42.6 & 38.4 & \cellcolor{Gray}39.5 & 33.3  & 35.7 & 33.3 & 34.5 & 41.6 \\
10$^{-3}$ & 46.5 & 41.8 & 35.0 & 38.2  & 34.0 & 41.3 & 31.6 & \cellcolor{Gray}\textbf{49.2} \\
10$^{-2}$ & 48.9 & 36.5 & 37.6 & \cellcolor{Gray}40.6  & 34.0 & 48.5 & 32.1 & 46.6 \\
10$^{-1}$ & 50.2 & \cellcolor{Gray}42.3 & 36.8 & 35.5  & 34.9 & \cellcolor{Gray}49.1 & \cellcolor{Gray}36.9 & 46.6 \\
0$^{-0} $ & \cellcolor{Gray}\textbf{55.0} & 38.3 & 37.8 & 31.1  & \cellcolor{Gray}38.9 & 38.3 & 36.2 & 46.5 \\
\bottomrule
\end{tabularx}
\vspace{-2.5mm}
\end{table*}

\subsection{Linguistic System}\label{saa}

In contrast to general acoustic modelling using the raw wave form as input, the extraction of semantics from text has been a well established research task for natural language processing (NLP) for some time now. 
Consequently a large variety of pretrained models have been made available freely. 
Specifically transformer architectures have recently raised the bar for many NLP tasks \cite{devlin2018bert}.

Instead of training a new model from scratch we therefore choose a pretrained BERT \cite{devlin2018bert} model provided by Wolf et al. \cite{Wolf2019HuggingFacesTS}.
The model has been trained for the German language on a large dataset composed of German Wikipedia dumps, legal texts and news articles.  

To extract feature vectors from the transcript of each narrative we first split it into separate sentences using the Spacy library \cite{Honnibal2017spacy2}. 
This allows us to classify semantic units separately and therefore achieve a finer granularity. 
The final assessment of the overall narrative is built by applying a majority voting mechanism over all classifications. 
For each sentence we apply tokenization to transfer them to the format that is expected by our pretrained model without any further preprocessing. 

The final extracted representations for each sample are then calculated by concatenating the mean and the max values for each sequence per output neuron. 
Like for the acoustic model we receive a one dimensional feature vector with the length of 1536.

\begin{figure}[]
  \includegraphics[width=\linewidth]{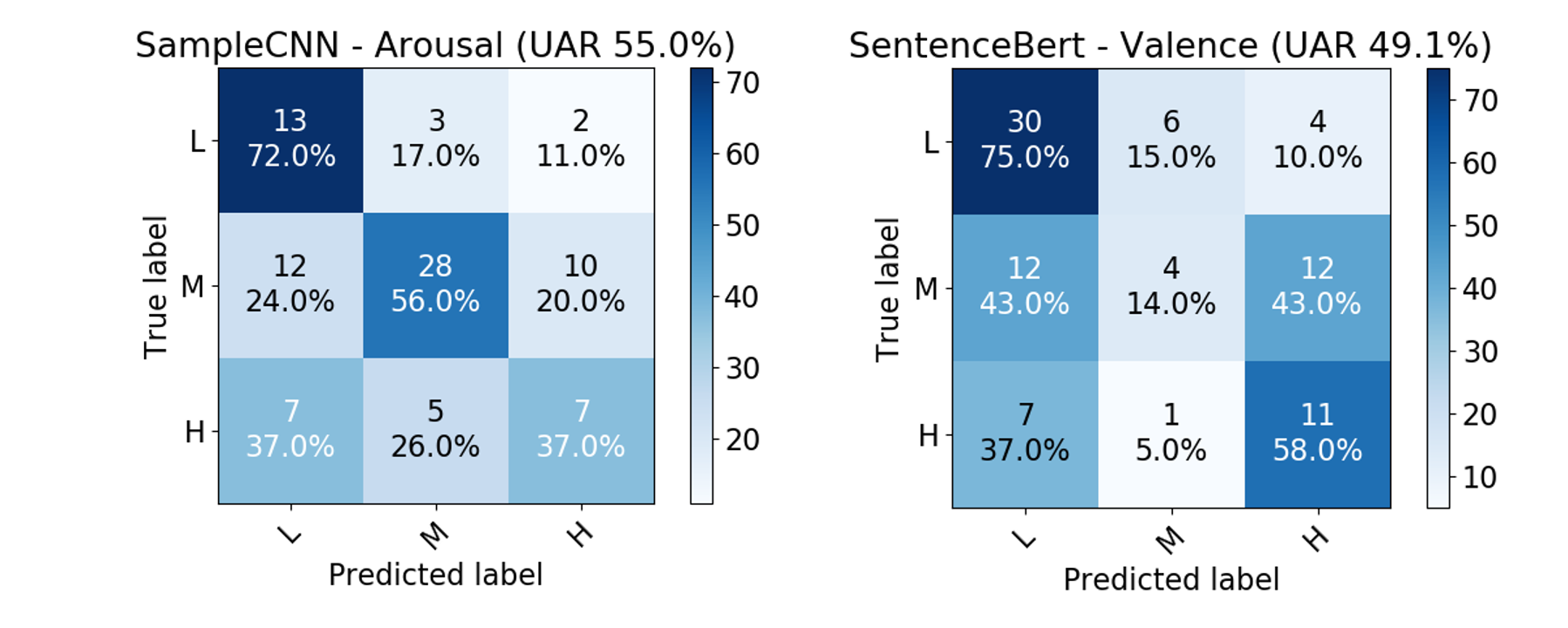}
  \caption{Confusion matrices for our best models on the development set.}
  \label{fig:cm}
\end{figure}

\section{Dataset}\label{datasets}

We run all our experiments on the \textit{Ulm State of Mind in Speech-elderly} dataset provided by the organizers of the challenge \cite{compare2020}.
The dataset consists of 87 elderly people telling negative and positive spontaneous stories.
Each narrative is labeled with respect to valence and arousal, created by calculating the mean of the speakers self-assessed emotional state and a post hoc annotation through experts.
The labels have been discretized to reflect the values \textit{low}, \textit{medium} and \textit{high}.
For each narrative the audio recording has been provided in the form of 5 second splits with a 16kHz sample rate, mono channel. 
Further more the challenge organizers provided a manually as well as an automatically created transcript for each narrative.
The data has been divided into three sets: Train, Development and Test, each consisting of the narratives from 29 speakers. 
Speech of one single speaker is not contained in multiple partitions.
Only the labels for Train and Development were made available to the participants of the challenge.
For more details please refer to \cite{compare2020}.

\section{Results}\label{results}

In the following we present the results of our experiments conducted for the Elderly Emotion Sub-Challenge. 
Performance will be reported in terms of Unweighted Average Recall (UAR), the official metric for this task in the ComParE Challenge. 
In order to keep the results comparable to the challenge baseline we are using both systems described in this paper only to extract features and perform the final classification with the system suggested by the authors of the challenge. 
That is, we train a SVM classifier with varying complexities (see \cite{compare2020}).
The final result for one narrative is calculated by a majority vote over all classified samples (5 second splits for audio and sentence wise for the transcript). 

In Table \ref{tab:results-dev} we compare our results to the performance of the two official baseline models on the development set. 
Similar to our own approaches, the two baseline systems can be divided into an acoustic model (\textit{DeepSpectrum}) and a linguistic model (\textit{LiFE}), which are extracting features from the input.
In contrast to our own acoustic model, the \textit{DeepSpectrum} approach operates in the frequency domain, which requires an additional, potentially lossy preprocessing step for domain transformation. 
Like our SentenceBert model, the LiFE baseline model also relies on pretrained BERT embeddings. 
The baseline approach, however, uses a recurrent neural network to learn a compressed version of those embeddings for each narrative.
Those compressions are specifically trained on the target corpus, which distinguishes the baseline approach from ours since we explicitly investigate the feasibility of generic features.  
For details please refer to \cite{compare2020}.

On the left part of the table the results are shown for each model with respect to the arousal task, while the right side shows the performance for each model on the valence task. 
The optimal training complexity for each model is highlighted grey and the best overall model for each task is printed in bold. 
For arousal we can observe that both our models outperform the baseline models, with the SampleCNN model being the clear winner in this category with a UAR of 55.0\%. 
In case of valence we can see that the SentenceBert model performs almost as good as the LiFE model, the best model in this category, with an UAR of 49.1\% compared to 49.2\%. 
Both the SampleCNN and the DeepSpectrum approaches are falling behind in performance. 

In addition to the results on the development set we report our best results on the test set in comparison to the official baseline. 
Our best models have been chosen by means of the highest unweighted average recall during training but before applying majority votes. 
The results are shown in Table \ref{tab:results-test}. 
For arousal the DeepSpectrum baseline system performs best in class with an UAR of 50.4\%.
On the other side our SentenceBert system achieves a considerably better result than the proposed baseline, with 5.7 percentage points improvement.

\section{Discussion}\label{discussion}
In this paper we have presented two neural network architectures that have been pretrained to learn meaningful feature representations from low level input data. 
The main question we seek to answer is whether the learned representations, that can be extracted by those networks, are feasible features for the task of automatic emotion recognition from narratives told by elderly people.
A closer look at the results on the development set (see Table \ref{tab:results-dev}) shows that the SampleCNN model, that operates on the raw audio waveform, performs considerably better for classifying arousal (UAR 55.0\%) than valence (UAR 38.9\%).
At the same time the transcript based features that have been extracted via the SentenceBert model are better suited for classifying valence (UAR 49.1\%) then arousal (UAR 42.3\%).
This observation should be expected since it is known that valence can be better modeled from semantic context while arousal can be better detected by the analysis of the paralinguistic properties. 
This finding is in line with the observation that has also been made by the authors of the challenge \cite{compare2020}.

Figure \ref{fig:cm} displays the confusion matrix for the SampleCNN and the SentenceBert model on the development set. 
The acoustic model on the left side shows a clear tendency towards a diagonal distribution within the matrix.
However, it is clearly visible that the model performs better for cases of lower and mid-range arousal than high arousal. 

On the other side our valence model performs best for the extreme cases of low and high valence but drops in performance for the middle class. 
During a manual inspection of random samples we found that instances that are labeled with high or low valence often contain distinct positive and negative keywords and phrases (e.g '\textit{happy}', '\textit{sad}', '\textit{positive} story'). 
This could potentially lead to improved recognition rates for those classes, when compared to the middle class.

Finally we would like to discuss the performance on the challenge test dataset. 
When directly comparing the linguistic and acoustic models to their respective baseline counterparts, we can see that our proposed approaches are performing almost equal or better in all cases on the development set. 
On the test set however only the SentenceBert model manages to outperform the baseline, while our SampleCNN model falls clearly behind.
A possible explanation for this considerable drop in performance between the development set and the test set regarding our acoustic model could be that there is a certain discrepancy within the domain or label distribution between the two sets.
This seems reasonable since the baseline classifiers have been chosen with respect to their highest performance on the test set and are showing a similar inverse proportional behaviour towards the development set. 
It is also noteworthy that all our models were trained only on the train set, to keep the results comparable. 
However, the official baseline models are trained on the combination of Train and Development which might result in better generalization capabilities of the model. 

\begin{table}[]
\caption{Results on the test set.}
\label{tab:results-test}
\begin{tabular}{cc|cc}
\toprule
\multicolumn{4}{c}{\emph{UAR [\%] Test (A/V)}} \\ \midrule 
SampleCNN & SentenceBert & DeepSpectrum & LiFE \\ \midrule
39.5 / - & - / \textbf{54.7} & \textbf{50.4} / 40.3 & 44.0 / 49.0\\

\bottomrule
\end{tabular}
\end{table}

\section{Conclusion}\label{conclusion}
The given task of elderly emotion classification turns out to be a difficult one. 
We showed that our method of applying previously learned feature representations to solve the problem performs equal or better than the baseline systems for all tasks on the development set.
However, for the test set we achieved mixed performance where we clearly outperformed the baseline for valence but falling short on arousal. 

We conclude that our proposed approach to extract feature representations from raw input data rivals methods that have been trained to extract features for a specific dataset as well as handcrafted features. 
This not only eliminates the time-consuming feature engineering phase, but also enables faster and more efficient training routines.

\section{Acknowledgements}\label{Acknowledgements}
This work has received funding from the European Union under grant agreement No 847926 Mindbot and 856879 Present. 

\bibliographystyle{IEEEtran}

\bibliography{bib}

\begin{thebibliography}{10}
\providecommand{\url}[1]{#1}
\csname url@samestyle\endcsname
\providecommand{\newblock}{\relax}
\providecommand{\bibinfo}[2]{#2}
\providecommand{\BIBentrySTDinterwordspacing}{\spaceskip=0pt\relax}
\providecommand{\BIBentryALTinterwordstretchfactor}{4}
\providecommand{\BIBentryALTinterwordspacing}{\spaceskip=\fontdimen2\font plus
\BIBentryALTinterwordstretchfactor\fontdimen3\font minus
  \fontdimen4\font\relax}
\providecommand{\BIBforeignlanguage}[2]{{%
\expandafter\ifx\csname l@#1\endcsname\relax
\typeout{** WARNING: IEEEtran.bst: No hyphenation pattern has been}%
\typeout{** loaded for the language `#1'. Using the pattern for}%
\typeout{** the default language instead.}%
\else
\language=\csname l@#1\endcsname
\fi
#2}}
\providecommand{\BIBdecl}{\relax}
\BIBdecl

\bibitem{weninger2013acoustics}
F.~Weninger, F.~Eyben, B.~W. Schuller, M.~Mortillaro, and K.~R. Scherer, ``On
  the acoustics of emotion in audio: what speech, music, and sound have in
  common,'' \emph{Frontiers in psychology}, vol.~4, p. 292, 2013.

\bibitem{mohammad2013crowdsourcing}
S.~M. Mohammad and P.~D. Turney, ``Crowdsourcing a word--emotion association
  lexicon,'' \emph{Computational Intelligence}, vol.~29, no.~3, pp. 436--465,
  2013.

\bibitem{wagner2018handcrafted}
\BIBentryALTinterwordspacing
J.~Wagner, D.~Schiller, A.~Seiderer, and E.~Andr{\'{e}}, ``Deep learning in
  paralinguistic recognition tasks: Are hand-crafted features still relevant?''
  in \emph{Interspeech 2018, 19th Annual Conference of the International Speech
  Communication Association, Hyderabad, India, 2-6 September 2018},
  B.~Yegnanarayana, Ed.\hskip 1em plus 0.5em minus 0.4em\relax {ISCA}, 2018,
  pp. 147--151. [Online]. Available:
  \url{https://doi.org/10.21437/Interspeech.2018-1238}
\BIBentrySTDinterwordspacing

\bibitem{trigeorgis2016adieu}
G.~Trigeorgis, F.~Ringeval, R.~Brueckner, E.~Marchi, M.~A. Nicolaou, B.~W.
  Schuller, and S.~Zafeiriou, ``Adieu features? end-to-end speech emotion
  recognition using a deep convolutional recurrent network,'' in
  \emph{Proceedings of the {IEEE} International Conference on Acoustics, Speech
  and Signal Processing, {ICASSP}}, Shanghai, China, 2016, pp. 5200--5204.

\bibitem{polignano2019comparison}
M.~Polignano, P.~Basile, M.~de~Gemmis, and G.~Semeraro, ``A comparison of
  word-embeddings in emotion detection from text using bilstm, cnn and
  self-attention,'' in \emph{Adjunct Publication of the 27th Conference on User
  Modeling, Adaptation and Personalization}, 2019, pp. 63--68.

\bibitem{compare2020}
B.~W. Schuller, A.~Batliner, C.~Bergler, E.-M. Messner, A.~Hamilton,
  S.~Amiriparian, A.~Baird, G.~Rizos, M.~Schmitt, L.~Stappen, H.~Baumeister,
  A.~D. MacIntyre, and S.~Hantke, ``The {INTERSPEECH} 2020 {C}omputational
  {P}aralinguistics {C}hallenge: {E}lderly emotion, {B}reathing \& {M}asks,''
  in \emph{Proceedings of Interspeech}, Shanghai, China, September 2020, p. 5
  pages, to appear.

\bibitem{rajan2019conflictnet}
V.~Rajan, A.~Brutti, and A.~Cavallaro, ``Conflictnet: End-to-end learning for
  speech-based conflict intensity estimation,'' \emph{IEEE Signal Processing
  Letters}, vol.~26, no.~11, pp. 1668--1672, 2019.

\bibitem{park2019specaugment}
D.~S. Park, W.~Chan, Y.~Zhang, C.-C. Chiu, B.~Zoph, E.~D. Cubuk, and Q.~V. Le,
  ``Specaugment: A simple data augmentation method for automatic speech
  recognition,'' \emph{arXiv preprint arXiv:1904.08779}, 2019.

\bibitem{perez2017effectiveness}
L.~Perez and J.~Wang, ``The effectiveness of data augmentation in image
  classification using deep learning,'' \emph{arXiv preprint arXiv:1712.04621},
  2017.

\bibitem{fadaee2017data}
M.~Fadaee, A.~Bisazza, and C.~Monz, ``Data augmentation for low-resource neural
  machine translation,'' \emph{arXiv preprint arXiv:1705.00440}, 2017.

\bibitem{panayotov2015librispeech}
V.~Panayotov, G.~Chen, D.~Povey, and S.~Khudanpur, ``Librispeech: an asr corpus
  based on public domain audio books,'' in \emph{Acoustics, Speech and Signal
  Processing (ICASSP), 2015 IEEE International Conference on}.\hskip 1em plus
  0.5em minus 0.4em\relax IEEE, 2015, pp. 5206--5210.

\bibitem{sharif2014cnn}
A.~Sharif~Razavian, H.~Azizpour, J.~Sullivan, and S.~Carlsson, ``Cnn features
  off-the-shelf: an astounding baseline for recognition,'' in \emph{Proceedings
  of the IEEE conference on computer vision and pattern recognition workshops},
  2014, pp. 806--813.

\bibitem{gwardys2014deep}
G.~Gwardys and D.~M. Grzywczak, ``Deep image features in music information
  retrieval,'' \emph{International Journal of Electronics and
  Telecommunications}, vol.~60, no.~4, pp. 321--326, 2014.

\bibitem{yenigalla2018speech}
P.~Yenigalla, A.~Kumar, S.~Tripathi, C.~Singh, S.~Kar, and J.~Vepa, ``Speech
  emotion recognition using spectrogram \& phoneme embedding.'' in
  \emph{Interspeech}, 2018, pp. 3688--3692.

\bibitem{fedotov2019time}
D.~Fedotov, B.~Kim, A.~Karpov, and W.~Minker, ``Time-continuous emotion
  recognition using spectrogram based cnn-rnn modelling,'' in
  \emph{International Conference on Speech and Computer}.\hskip 1em plus 0.5em
  minus 0.4em\relax Springer, 2019, pp. 93--102.

\bibitem{karadougan2012combining}
S.~G. Karado{\u{g}}an and J.~Larsen, ``Combining semantic and acoustic features
  for valence and arousal recognition in speech,'' in \emph{2012 3rd
  International Workshop on Cognitive Information Processing (CIP)}.\hskip 1em
  plus 0.5em minus 0.4em\relax IEEE, 2012, pp. 1--6.

\bibitem{lee2018samplecnn}
J.~Lee, J.~Park, K.~L. Kim, and J.~Nam, ``Samplecnn: End-to-end deep
  convolutional neural networks using very small filters for music
  classification,'' \emph{Applied Sciences}, vol.~8, no.~1, p. 150, 2018.

\bibitem{dieleman2014end}
S.~Dieleman and B.~Schrauwen, ``End-to-end learning for music audio,'' in
  \emph{2014 IEEE International Conference on Acoustics, Speech and Signal
  Processing (ICASSP)}.\hskip 1em plus 0.5em minus 0.4em\relax IEEE, 2014, pp.
  6964--6968.

\bibitem{kim2019comparison}
T.~Kim, J.~Lee, and J.~Nam, ``Comparison and analysis of samplecnn
  architectures for audio classification,'' \emph{IEEE Journal of Selected
  Topics in Signal Processing}, vol.~13, no.~2, pp. 285--297, 2019.

\bibitem{devlin2018bert}
J.~Devlin, M.-W. Chang, K.~Lee, and K.~Toutanova, ``Bert: Pre-training of deep
  bidirectional transformers for language understanding,'' \emph{arXiv preprint
  arXiv:1810.04805}, 2018.

\bibitem{Wolf2019HuggingFacesTS}
T.~Wolf, L.~Debut, V.~Sanh, J.~Chaumond, C.~Delangue, A.~Moi, P.~Cistac,
  T.~Rault, R.~Louf, M.~Funtowicz, and J.~Brew, ``Huggingface's transformers:
  State-of-the-art natural language processing,'' \emph{ArXiv}, vol.
  abs/1910.03771, 2019.

\bibitem{Honnibal2017spacy2}
M.~Honnibal and I.~Montani, ``{spaCy 2}: Natural language understanding with
  {B}loom embeddings, convolutional neural networks and incremental parsing,''
  2017, to appear.

\end{thebibliography}

\end{document}